\documentclass[onecolumn,aps,prb]{revtex4}
\usepackage{epsf,graphicx}
                                                                                
\begin{document}

\draft

\title{Finite size effects on calorimetric cooperativity
of two-state proteins}

\author{Mai Suan Li$^1$, D. K. Klimov$^2$ and D. Thirumalai$^2$}
                                                                                
\address{$^1$Institute of Physics, Polish Academy of Sciences,
Al. Lotnikow 32/46, 02-668 Warsaw, Poland\\
$^2$Department of Chemistry and Biochemistry and Institute for Physical
Science and Technology, University of Maryland, College Park, MD 20742 }

\begin{abstract}
Finite size effects on the calorimetric cooperatity of the
folding-unfolding transition in two-state proteins are considered
using the Go lattice models with and without side chains.
We show that for models without
side chains a dimensionless measure of 
calorimetric cooperativity $\kappa _2$ defined as
the ratio of the van't Hoff 
to calorimetric enthalpy 
does not depend on the number
of amino acids $N$. The
average value $\overline{\kappa_2}
\approx \frac{3}{4}$ is lower than the experimental
value $\kappa_2 \approx 1$. For models with side chains $\kappa _2$
approaches unity as $\kappa _2 \sim N^{\mu}$, where 
$\mu \approx 0.17$. Above the critical
chain length
$N_c \approx 135$ these models can mimic the truly all-or-non
folding-unfolding transition.
\end{abstract}

\maketitle

\section{Introduction}
Single domain globular proteins, which are finite sized objects,
undergo remarkably cooperative transitions from an 
ensemble of unfolded states to well ordered folded (or native) states
as the temperature is lowered \cite{Finkel_book}. 
In many cases, the transition to the native state takes place
in an 
apparent two-state manner, i.e. the only detectable species are the native 
(more precisely, the ensemble of conformations belonging to
the native basin of attraction \cite{Li99})
or unfolded states \cite{Poland}.
In order to characterize the two-state folding one can use the
dimensionless quantity
$\kappa _2$ \cite{Chan00PRL}
\begin{eqnarray}
\kappa _2\; \, = \; \, \Delta H_{vh}/\Delta H_{cal} \; ,
\label{VantHoff_eq}
\end{eqnarray}
where 
$\Delta H_{vh} \;  = \; 2T_{max}\sqrt{k_B C_P(T_{max})}$
and $\Delta H_{cal} \; = \;  \int_0^{\infty} C_P(T)dT$,
are the van't Hoff and the calorimetric enthalpy, respectively,
$C_P(T)$ is the specific heat.
$\kappa _2$ may be considered as a measure of the calorimetric
cooperativity.
Since real globular
proteins have $\kappa _2$ very close to unity (chymotrypsin
inhibitor 2 is a prime example \cite{Jackson91})
it was proposed that \cite{Chan04} $\kappa _2 \approx 1$ can serve
as one of requirements for realistic models of proteins.
There are technical problems in evaluating $\kappa_2$ using
experiments or
computations. Inadequate treatment of baseline subtractions in
$C_P(T)$ obscures estimates of $\kappa_2$. As a result it is
possible
that even sequences with $\kappa_2 \approx 1$ may not clearly
be
two-state folders. Nevertheless, $\kappa_2$ or related measures
have often
been used as a measure of calorimetric cooperativity.

In series of works \cite{Chan00PRL,Chan00,Chan03,Chan04} Chan 
{\em et al.} have shown that the calorimetric criterion is
difficult to satisfy theoretically.
Even Go models \cite{Go83} which are more cooperative than
others (2-letter, 3-letter and 20-letter models) have $\kappa _2$
notably smaller than 1. The studies of the Chan group are
limited to few sequences and it remains, therefore,
unclear if the Go modeling can meet the calorimetric
requirement. One of our goals is to try to solve this problem by
carrying out comprehensive simulations 
of lattice Go models.

Another dimensionless measure of thermodynamic cooperativity
is $\Omega _c$ defined as follows \cite{Camacho93PNAS}
\begin{equation}
\Omega _c \; = \;  \; \frac{T_F^2}{\Delta T}
\biggl(\frac{d<\chi>}{dT}\biggr)_{T=T_F}.
\label{coop_eq}
\end{equation}
Here $\chi$ is the structural overlap with the native state
and it can be identified as the probability of occupation of the
native basin of attraction, $T_F$ is the folding temperature
and $\Delta T$ is the transition width \cite{Li04POLY}.
$\Omega _c$ may be referred to as the
structural cooperativity.
Recently, we have shown that \cite{Li04}
it grows with the
chain length as $\Omega _c \sim N^{\zeta}$, where
the universal exponent
$\zeta \approx 2.22$. This result is supported by experimental
data collected for 32 two-state wild type proteins and by 
simulations for lattice models.
The main goal of this paper is to 
consider the finite
size effects on $\kappa _2$ of two-state folders with the help
of lattice Go models and Monte Carlo simulations.
From the definition of $\kappa_2$ it follows that it should be
independent
of $N$ because both $\Delta H_{vh}$ and $\Delta H_{cal}$ are extensive
variables. However, the approach to the asymptotic behavior is
unclear.

We have studied two classes of models:
lattice models  without side chains (LM) and lattice models
with side chain (LMSC). For the first class, in accord with
experiments,
$\kappa_2$ was found to be scale-invariant at least up
to $N \le 80$. However, for 78 sequences studied
their average value
 $\overline{\kappa_2} \approx
\frac{3}{4}$ which is clearly smaller unity.
Thus, in agreement with the previous results
\cite{Chan00PRL,Chan00,Chan03,Chan04}, Go LMs
do not satisfy the proteinlike cooperativity principle
although they are minimally frustrated.

For Go LMSCs
 we have found that $\kappa _2$
scales with $N$ as
\begin{equation}
 \kappa _2 \; \sim \;  N^{\mu}
\label{scaling_eq}
\end{equation}
 before reaching
the maximal value 1 at  the critical value $N_c \approx 135$.
Here exponent $\mu = 0.17 \pm 0.02$.
These results
suggest that $\kappa _2$ becomes scale-invariant for
$N \gtrsim N_c$ and the LMSCs can meet the strict calorimetric
cooperativity criterion only for this range of system sizes.
If one assumes that the all-or-non folding takes place at
$\kappa_2 \gtrsim 0.9$ then the critical value $N_c$ is reduced
to $N^{\ast} = 70$ (see below). 
In this case the LMSC with $N \gtrsim N^{\ast}$ can capture the
calorimetric behavior of two-state proteins.

\begin{figure}
\epsfxsize=4in
\centerline{\epsffile{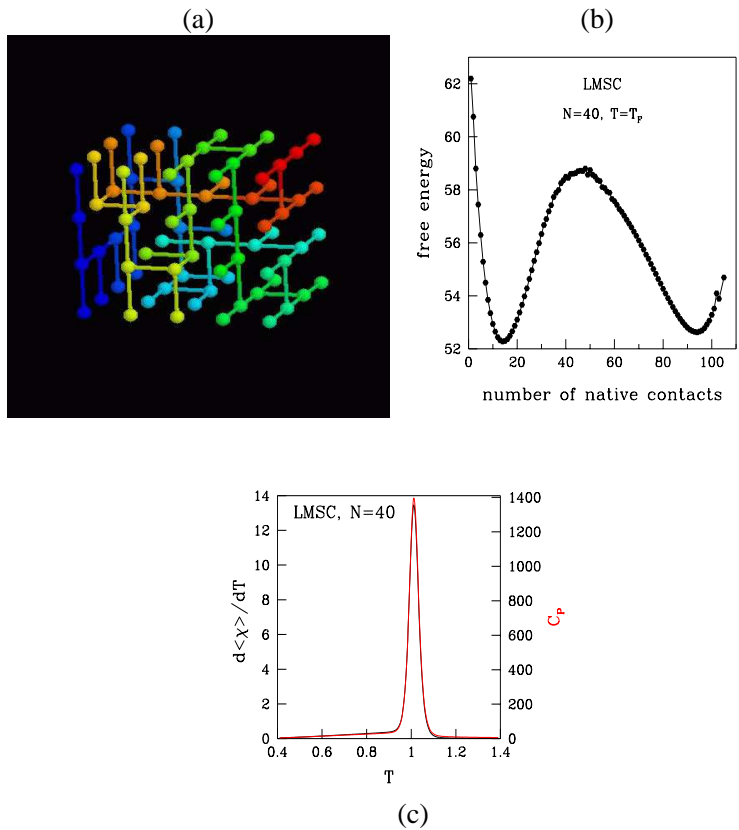}}
\vspace{0.1in}
\caption{(a) Typical native conformation of $N=40$ of the LMSC.
The BB and SC beads occupy sites of the compact
$4\times 4\times 5$ lattice. (b) Dependence
of the free energy (measured in $k_BT$) obtained for the sequence
 whose the native conformation is shown in a)
on the number of native contacts at $T=T_F$. Since the free energy
has
only one local maximum at the transition state
this sequence is a two-state folder.
(c) Temperature dependence of $d<\chi >/dT$ (black) and
$C_P$ (red, right-hand scale) for the sequence whose
 the native conformation is
shown in a).
}
\label{sc40_s1_fig}
\end{figure}

\section{Models and method}

In the coarse grained representation of LM each amino acid
is represented as a single bead confined to the vertices of a cubic
lattice \cite{Dill95ProtSci}. The LMSC is also modeled on
 a cubic
lattice by a backbone (BB) sequence of
$N$ beads, to which a "side" bead, representing a side chain,
is attached.
The peptide bond and the $\alpha$-carbon
are given by a single bead and
the system has in total 2$N$ beads.
Self-avoidance is imposed,
i.e. any backbone and side beads cannot occupy the same lattice
site more than once.

 In the LMSC
the energy of a conformation is  \cite{KlimThirum98FD,Li02JPC}
\begin{eqnarray}
E \; = \;  \epsilon _{bb} \sum_{i=1,j>i+1}^{N} \,
 \delta _{r_{ij}^{bb},a}
+ \epsilon _{bs} \sum_{i=1,j\neq i}^{N} \, \delta _{r_{ij}^{bs},a}
+ \epsilon _{ss} \sum_{i=1,j>i}^{N} \, \delta _{r_{ij}^{ss},a} \; ,
\label{energy_eq}
\end{eqnarray}
where $\epsilon _{bb}, \epsilon _{bs}$ and $\epsilon _{ss}$ are
BB-BB,
BB-SC and SC-SC contact energies.
$r_{ij}^{bb}, r_{ij}^{bs}$ and $r_{ij}^{ss}$ are
the distances
between the $i^{th}$ and $j^{th}$ residues for the  BB-BB,
BB-SC and SC-SC pairs, respectively, $a$ is lattice spacing.
Energies $\epsilon _{bb}, \epsilon _{bs}$
and $\epsilon _{ss}$ are chosen to be -1 for native contacts
and 0 for non-native ones.  For the LM the energy
in Eq. (\ref{energy_eq}) has only the BB term.

The specific heat in Eq. (\ref{VantHoff_eq}) is defined as 
the energy fluctuation. For LMSC
the overlap function $\chi$ 
is defined as
\begin{eqnarray}
\chi = \frac{1}{2N^{2}-3N+1} \Big[
\sum_{i < j} \delta(r^{ss}_{ij} - r^{ss, N}_{ij}) +
\sum_{i < j + 1} \delta(r^{bb}_{ij} - r^{bb, N}_{ij}) + \nonumber\\
\sum_{i \neq j} \delta(r^{bs}_{ij} - r^{bs, N}_{ij}) \Big],
\label{chi_eq}
\end{eqnarray}
here the upper script $N$ refers to the native state
and factor $2N^{2}-3N+1$ ensures that $\chi=1$ in the
native conformation. The last equation with only the BB
term  is applied to the LMs. 

The Monte Carlo simulations were carried out
using the move
set MS3 \cite{Betancourt,Li02JPC,Li02CPC} which involves single, double and triple
bead moves. Because this move set involves multiparticle updates it is
much more efficient compared to the standard
move set \cite{Hilhorst75}.
The thermodynamic properties are calculated using the multiple
histogram method \cite{Ferrenberg89}. Sequences are selected as
two-state folders if their free energy plotted against the number
of native contacts has two well-defined minima.

\begin{figure}
\epsfxsize=4in
\centerline{\epsffile{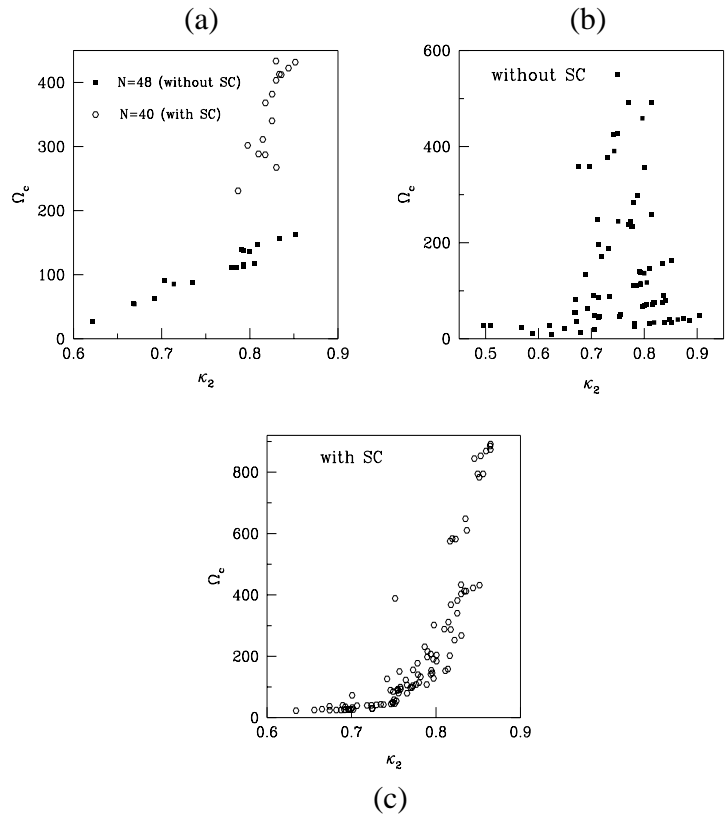}}
\vspace{0.1in}
\caption{The dependence of $\Omega _c$ on $\kappa_2$ for
$N=48$ LMs (solid squares, 18 sequences) and $N=40$ LMSC
(open hexagons, 15 sequences) (a), for all $N$ LMs (b)
and for all $N$ LMSCs (c).
For LMs we have studied $N=27 (17), 36 (17), 48 (18),
64 (15)$ and 80 (11) and for LMSCs - $N=$ 18 (30),
24 (18), 32 (20), 40 (15) and 50 (15). Numbers of studied
sequences are indicated in  parenthesis.
}
\label{bt_vhove_fig}
\end{figure}

\section{Results}

Fig. \ref{sc40_s1_fig}a shows the typical native conformation
of the $N=40$ LMSC sequence. The free energy is calculated as
a function of the
number of native contacts, which is  treated  as
an approximate reaction
coordinate for Go models, and the corresponding results
obtained at $T=T_F$ are shown
in Fig. \ref{sc40_s1_fig}b. Since the free energy profile has
only one local maximum  
located at the transition state this sequence is a two-state
folder. Clearly, for Go models the peaks of $C_P$ and $d<\chi>/dT$ coincide 
(Fig. \ref{sc40_s1_fig}c).

Fig. \ref{bt_vhove_fig}a shows the structural cooperativity
against the calorimetric one for a given value of $N$. As expected,
$\Omega_c$ grows with $\kappa_2$ for both LMs and LMSCs. 
However, the relation between these quantities becomes
non-trivial if we combine the results for all values of $N$
(Fig. \ref{bt_vhove_fig}b and Fig. \ref{bt_vhove_fig}c).
 The correlation remains strong
for LMSCs but surprisingly it almost vanishes for LMs. 
It is not clear if the absence of correlation for the LMs is 
intrinsic or it is merely an artifact of the limited set of data.
Clarification of this point requires further investigation.
From all sequences 176 sequences studied
(78 LM sequences and 98 LMSC ones) 10 sequences
have $\kappa_2 \gtrsim 0.85$ and only one sequence which
has $\kappa_2 \approx 0.9$ nearly satisfies the calorimetric
cooperativity principle. 

Since $\kappa_2$
of the LMs is not sensitive to $N$ we can calculate its averaged
value over the whole data set (78 sequences) and obtain 
$\overline{\kappa_2} \approx \frac{3}{4}$ which is notably
smaller than unity. Thus our results,
 which are in accord with Kaya and Chan, also
 suggest that it is hard
to meet the calorimetric criterion for Go LMs for any
chain length. Using the
relation $\kappa_2 = \sqrt{1-4(T_G/T_F)^2}$
derived from the random energy model \cite{Chan00,Chan00a,Chan04},
where $T_G$ is interpreted as the temperature below which
folding kinetics is
dominated by trapping mechanisms \cite{Onuchic97ARPC}, we obtain
$\frac{T_F}{T_G}=\frac{8}{\sqrt{7}} \approx 3$. This value is 
far below the proposed
$\frac{T_F}{T_G}=4.6$ \cite{Chan04} required for the two-state 
melting with $\kappa_2=0.9$ but higher than, say, 
$\frac{T_F}{T_G}=1.6$ for three-letter models \cite{Chan00}.

\begin{figure}
\epsfxsize=4in
\centerline{\epsffile{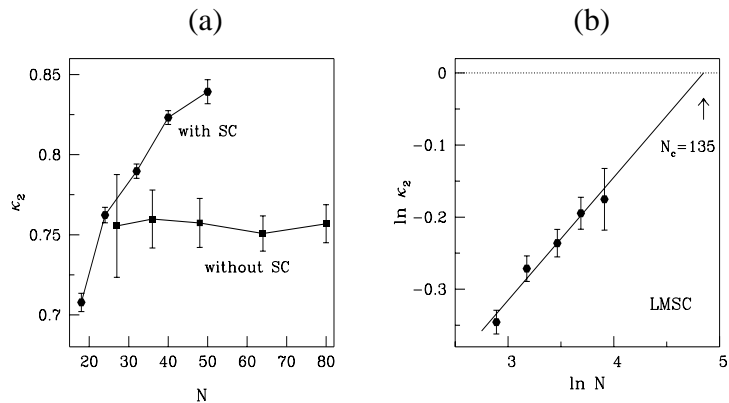}}
\vspace{0.2in}
\caption{(a) Dependence of $\kappa_2$ on $N$ for LMs
(solid squares) and LMSCs (solid hexagons).
The sequences are the same as in Fig. \ref{bt_vhove_fig}.
 (b) The same
as for LMSCs in a) but data are shown in the log-log plot.
The dotted line refers to $\kappa_2=1$.
The solid straight line is linear fit $y = -0.809 + 0.165 x$
(the correlation coefficient is 0.96).
It crosses the $\kappa_2=1$ line at the critical value
$N_c=135$. 
}
\label{scaling_fig}
\end{figure}

The difference in the scaling behavior of LMs and LMSCs is
clearly seen
in Fig. \ref{scaling_fig}a where the size effect is visible only
for sequences with SC. From the log-log plot 
(Fig. \ref{scaling_fig}b) we obtain exponent 
$\mu=0.17 \pm 0.02$. Interpolating our results to
$\kappa_2=1$ we find the critical length $N_c \approx 135$
above which  LMSCs always satisfy the calorimetric cooperativity
requirement. If we assume that the transition is two-state if
$\kappa_2 \gtrsim 0.9$ then the calorimetric cooperativity is
satisfied for $N \gtrsim N^{\ast}$, where $N^{\ast} \approx 70$.  

\section{Conclusion}

We have shown that for a given system size the structural
cooperativity correlates with
the calorimetric one. The
scaling of the calorimetric cooperativity has been examined
for lattice two-state Go models of proteins. The LMs 
superficially mimic
experiments in the sense that $\kappa_2$ is almost insensitive
to the system sizes. However, they are not able to
reproduce
the experimental value $\kappa_2 \approx 1$. The rate of
success for
designing a Go LM which have $\kappa_2 \gtrsim 0.9$ is rather
low (about $1\%$).
The lack of scaling of LM folding cooperativity with
chain length prevents these models to describe the cooperativity
of
wild-type proteins. This appears to be an inherent deficiency of
LM
without side chains.

For the Go LMSCs  $\kappa_2$
depends on the system size up to the critical size
$N_c$ above which the full requirement of the
calorimetric cooperativity is satisfied. Their advantage
is that the criterion 
$\kappa_2 \gtrsim 0.9$ may be satisfied for relatively
small globular proteins ($N \sim N^{\ast}=70$). 
Our study shows that incorporation of side chains in protein LM
represents a crucial modification, which makes LMSC protein-like.

It should be noted that we have considered the pairwise
interaction (\ref{energy_eq}) for Go models and it may be
the reason why the calorimetric criterion is hard to fulfill
even for LMSCs.
The  multiparticle
interactions may be required to quantitatively
describe cooperativity seen in proteins
\cite{Levitt97ProtSci,Chan03}.


This work was
supported by the KBN grant  No 1P03B01827
and the National Science Foundation grant
(NSF CHE-0209340). MSL thanks H. S. Chan for providing
Ref. \onlinecite{Chan04}.

\end{document}